\documentclass{amsproc}

\newtheorem{theorem}{Theorem}[section]

\newtheorem{quest}[theorem]{Question}

\theoremstyle{definition}

\theoremstyle{remark}

\numberwithin{equation}{section}

\newcommand{\beq}{\begin{equation}}
\newcommand{\eeq}{\end{equation}}
\newcommand{\ba}{\begin{array}}
\newcommand{\ea}{\end{array}}
\newcommand{\bea}{\begin{eqnarray}}
\newcommand{\eea}{\end{eqnarray}}
\newcommand{\bean}{\begin{eqnarray*}}
\newcommand{\eean}{\end{eqnarray*}}
\newcommand{\eref}[1]{(\ref{#1})}

\newcommand{\comment}[1]{}

\begin{document}

\title{Spectral Covers, Integrality Conditions, and Heterotic/F-theory Duality}

\author{Lara B. Anderson}
\address{Department of Physics, Virginia Tech, Blacksburg, Virginia 24060}
\email{lara.anderson@vt.edu}


\date{April 20th, 2014.}



\begin{abstract}
In this work we review a systematic, algorithmic construction of dual heterotic/F-theory geometries corresponding to $4$-dimensional, ${\mathcal N} = 1$ supersymmetric compactifications. We look in detail at an exotic class of well-defined Calabi-Yau fourfolds for which the standard formulation of the duality map appears to fail, leading to dual heterotic geometry which appears naively incompatible with the spectral cover construction of vector bundles. In the simplest class of examples the F-theory background consists of a generically singular elliptically fibered Calabi-Yau fourfold with $E_7$ symmetry. The vector bundles arising in the corresponding heterotic theory appear to violate an integrality condition of an $SU(2)$ spectral cover. A possible resolution of this puzzle is explored by studying the most general form of the integrality condition. This leads to the geometric challenge of determining the Picard group of surfaces of general type. We take an important first step in this direction by computing the Hodge numbers of an explicit spectral surface and bounding the Picard number.
\end{abstract}

\maketitle

\vspace{-15pt}
.
\section{An algorithm construction of dual heterotic/F-theory geometry}\label{hetfscan}
\noindent  Compactifications of heterotic string theory and F-theory are believed to be dual -- that is to lead to the same effective low energy physics -- whenever the compactification geometries take the form \cite{Vafa-F-theory,Morrison-Vafa-I,Morrison-Vafa-II, Friedman-mw}
\beq\label{theduals}
\text{Heterotic on}~~\pi_{h}: X_n \stackrel{\mathbb{E}}{\longrightarrow} B_{n-1}~~\Leftrightarrow ~~\text{F-theory on}~~ \pi_{f}: Y_{n+1} \stackrel{K3}{\longrightarrow} B_{n-1}
\eeq
where the $K3$ fiber of $Y_{n+1}$ is itself elliptically fibered over a $\mathbb{P}^1$ base. The compatibility of these two fibrations leads to the observation that $\rho_{f}: Y_{n+1} \stackrel{\mathbb{E}}{\longrightarrow} {\mathcal B}_{n}$ and $\sigma_{f}: {\mathcal B}_n \stackrel{\mathbb{P}^1}{\longrightarrow} B_{n-1}$. In recent work \cite{Anderson:2014gla} this duality was used to systematically enumerate an interesting and finite class of string backgrounds and the properties of the associated $4$-dimensional effective theories. As given in \eref{theduals}, the choice of geometry in F-theory consists simply of a $K3$-fibered Calabi-Yau fourfold. For the $E_8 \times E_8$ heterotic string theory the background is determined by an elliptically fibered Calabi-Yau threefold equipped with a pair of poly-stable, holomorphic vector bundles, $V_i$ ($i=1,2$) on $X_3$ with structure groups, $H_i \subset E_8$. 

In \cite{Anderson:2014gla} a program was set out to systematically study the general properties and constraints of the dual effective theories and develop a general and algorithmic formalism to build consistent heterotic/F-theory backgrounds. With this goal in mind, the first step in constructing a pair of the form \eref{theduals} is the choice of a twofold base, $B_2$ appearing in both the heterotic and F-theory geometry. For all smooth threefolds, $X_3$, the possible choices for $B_2$ have been classified \cite{gross} (and $B_2$ must be a generalized del Pezzo surface). Furthermore, to explore and test general structure there is an important dataset of such manifolds consisting of $61,539$ toric surfaces systematically constructed by Morrison and Taylor \cite{Morrison:2012np,Kumar:2009ac,Morrison:2012js}. 

With these results in place it iso possible to begin to build the geometry of \eref{theduals} from the bottom up. In the Calabi-Yau fourfold geometry the next step is to choose a form for the $\mathbb{P}^1$-fibration, $\sigma_{f}: {\mathcal B}_3 \stackrel{\mathbb{P}^1}{\longrightarrow} B_{2}$. As described in Section \ref{het_fth}, this can be accomplished for non-degenerate fibrations by building ${\mathcal B}_3$ as a $\mathbb{P}^1$ bundle over $B_2$, parameterized by a ``twist": a $(1,1)$-form $T$ in $B_2$ (see \eref{eq:m}). In the heterotic theory this choice of twist corresponds to a piece of the heterotic vector bundle topology (more specifically, a component of the second Chern class $c_2(V)$) \cite{Friedman-mw}. In \cite{Anderson:2014gla} we established that given a twofold base $B_2$, the set of all possible twists is in fact bounded by the conditions imposed by $4$-dimensional $N=1$ supersymmetry. In the heterotic theory this appears through the condition of slope stability of the vector bundles $V_i$ and in F-theory by the condition that the generically singular fourfold $Y_4$ admits a smooth Calabi-Yau resolution. Finally it should be noted that since we require all fibrations to admit (exactly one) section, each elliptically fibered manifold is birationally equivalent to a Weierstrass model \cite{nakayama} (see \eref{fweier}). Thus, having chosen $B_2$ and constructed a $\mathbb{P}^1$-bundle ${\mathcal B}_3$, we have fully specified $X_3$ and $Y_4$.

With consistency conditions in place and a scheme for algorithmically constructing pairs as in \eref{theduals}, it remains to extract patterns and structure from the effective theories. Duality here provides a powerful tool to determine otherwise difficult to calculate information on both sides of the theory. While historically heterotic/F-theory duality has been used to determine the effective physics of the mysterious and non-lagrangian F-theory, in \cite{Anderson:2014gla} we also explored ways in which  the singularity structure of the F-theory fourfold could be used to determine non-trivial information about ${\mathcal M}_{\omega}(c(V))$ -- the moduli space of sheaves that are semi-stable with respect to the K\"ahler form $\omega$ with fixed total Chern class $c(V)$. Such information is hard won, since very few techniques exist to determine ${\mathcal M}_{\omega}(c(V))$ for sheaves/bundles over Calabi-Yau threefolds (or their associated higher-rank Donaldson-Thomas invariants). 

As one simple illustration of this correspondence, we note here that the presence of generic symmetries on singular Calabi-Yau fourfolds make it possible to derive correlations between the topology of a slope-stable heterotic vector bundle on a CY threefold and its structure group. Initial investigations of this nature were first undertaken in \cite{Rajesh:1998ik,Berglund-Mayr} who constructed ``lower bounds" on the second Chern class of a vector bundle with fixed structure group. In \cite{Anderson:2014gla}, we continue to explore the links between structure group and topology, exploring not only these lower bounds but also upper bounds as well (see Section 6 of \cite{Anderson:2014gla}).

{\small \begin{table}[h!]
\begin{center}
\begin{tabular}{|l|l|l|l|}\hline
Structure Group, $H$ &Topology & Structure Group, $H$ & Topology \\\hline
$SU(N)$ & $\eta \geq N \cdot c_1(B_2)$ &$E_8$ & $\eta \geq 5 \cdot c_1(B_2)$ \\\hline
$SO(7)$ & $\eta \geq 4 \cdot c_1(B_2)$  & $E_7$ &$\eta \geq \frac{14}{3} \cdot c_1(B_2)$ \\\hline
$SO(M)$ & $\eta \geq \frac{M}{2} \cdot c_1(B_2)$ &$E_6$ &$\eta \geq \frac{9}{2} \cdot c_1(B_2)$\\\hline
$Sp(K)$ & $\eta \geq 2K \cdot c_1(B_2)$  &$G_2$ &$\eta \geq \frac{7}{2} \cdot c_1(B_2)$\\\hline
& & $F_4$ & $\eta \geq \frac{7}{2} \cdot c_1(B_2)$ \\ \hline
\end{tabular}\label{corr_top}
\caption{\it\small Constraints linking the topology, $\eta=c_2(V)|_{B_2}$, of an $H$-bundle $V$ and its structure group on an elliptically fibered CY threefold, $\pi_h: X_3 \to B_2$. \cite{Rajesh:1998ik,Berglund-Mayr}.}
\end{center}
\end{table}}
Systematic patterns such as those shown in Table \ref{corr_top} are of use in string phenomenology (for example they could simplify recent algorithmic searches for heterotic Standard Models carried such as those carried out in \cite{Anderson:2009mh,Anderson:2008uw,Anderson:2007nc,Anderson:2011ns,Anderson:2012yf,Anderson:2013xka}). In order to fully understand such patterns though, it is necessary to complete the geometric ``dictionary" which matches heterotic/F-theory geometry. This includes the inclusion of $G$-flux in the F-theory background and an understanding of the zero-locus of the induced Gukov-Vafa-Witten superpotential \cite{Gukov:1999ya}. In this context the quantization conditions on flux and the corresponding constraints in the heterotic theory become particularly important. Indeed, as described in detail \cite{Anderson:2014gla}, in our systematic search, we find many geometries which appear mysterious from the point of view of these commonly assumed integrality conditions.

In the following sections we will review the standard construction of heterotic/F-theory dual pairs. In its most explicit form, the duality map dependences on a particular method of constructing Mumford poly-stable vector bundles -- namely, the {\it spectral cover construction} \cite{Friedman-mw}. In recent work \cite{Anderson:2014gla, Anderson_stringmath} it has been observed that many apparently consistent F-theory fourfolds have topology which appears to be inconsistent with a naive construction of spectral cover bundles. We will explore this discrepancy further in concrete examples in the following Sections.

Out of the dataset generated in \cite{Anderson:2014gla}, we consider one of the simplest examples of such an exotic heterotic/F-theory dual pair. In particular, we explore the so-called ``integrality" condition on the spectral data (see \eref{mod2_cond}) and set out to determine whether it is really correct/necessary as frequently implemented in the literature. In addition, we lay out the necessary geometric questions that must be addressed if this criterion is to be refined or improved. We will argue that in many cases the surface forming the $SU(2)$ spectral cover can have a larger Picard group than is generically assumed and that the heterotic bundle can in fact be described by a consistent spectral cover pair $({\mathcal S}, {\mathcal L}_{S})$, consisting of a $2$-sheeted cover $\pi_{\mathcal{S}}: {\mathcal S} \to B_2$ and a line bundle over it ${\mathcal L}_{S}$ over it. We begin with a brief review of heterotic/F-theory duality in $4$-dimensions to set the stage for these investigations.

\section{Heterotic/F-theory Duality in $4$-dimensions}\label{het_fth}
In this section we will provide a rough outline of the geometric correspondence that arises in heterotic and F-theory dual pairs. Many excellent reviews exist in the literature and we refer the reader to classic sources such as \cite{Friedman-mw, Curio:1998bva} and modern summaries such as \cite{Anderson:2014gla} for a more complete treatment. In recent work, \cite{Anderson:2014gla} a constructive algorithm was developed to consistently build and enumerate dual heterotic/F-theory geometries. As a tractable starting point for that work, heterotic backgrounds were considered consisting of a smooth elliptically fibered Calabi-Yau threefold $X_3$ (with a single section\footnote{For geometries without section and some of the physics of these more general genus-1 fibrations see recent progress in \cite{Braun:2014oya,Morrison:2014era,Anderson:2014yva}.}) over a base $B_2$, together with two holomorphic, Mumford poly-stable vector bundles \cite{GSW}.  In such
cases, the dual F-theory compactification geometry can be built beginning with a rationally fibered threefold base ${\mathcal B}_3$ that is a $\mathbb{P}^1$ bundle
over $B_2$ (the same surface used to define the heterotic Calabi-Yau threefold).  The F-theory compactification space is then an elliptically and K3-fibered fourfold, $\rho_{f}: Y_4 \to {\mathcal B}_3$. Following \cite{Friedman-mw}, without loss of generality, the non-degenerate $\mathbb{P}^1$-fibered base (${\mathcal B}_3$) 
can be defined as a $\mathbb{P}^1$ bundle through the projectivization of a sum of two line bundles
\begin{equation}
{\mathcal B}_3 = \mathbb{P} ({\mathcal O} \oplus{\mathcal L}) \,,
\label{eq:m}
\end{equation}
where ${\mathcal L}$ is a general line bundle  on the base $B_2$.
Over ${\mathcal B}_3$, the classes $R = c_1 ({\mathcal O} (1)), T = c_1 ({\mathcal
  L})$, can be defined, where ${\mathcal O} (1)$ is a bundle that restricts to the usual
${\mathcal O} (1)$ on each $\mathbb{P}^1$ fiber.  The $\mathbb{P}^1$ fibration is equipped with sections $\Sigma_-$
and $\Sigma_+ = \Sigma_-+ T$ of  ${\mathcal B}_3$ satisfying $\Sigma_- \cdot \Sigma_+ = 0$, corresponding to the relation $R
(R + T) = 0$ in cohomology.

Finally, then the fourfold itself can be described in Weierstrass form as
\beq\label{fweier}
y^2=x^3 +f x +g
\eeq
where $y,x$ are (affine) coordinates along the elliptic fiber and $f \in H^0({\mathcal B}_3,K_{3}^{-4})$, $g \in H^0(\mathcal{B}_3, K_3^{-6})$. As usual the position of singular fibers is encoded in the discriminant locus, $\Delta = 4f^3 + 27g^2$.

For this choice of an F-theory model on $Y_4$ and a heterotic theory on $X_3$, it is now possible to begin by matching topology \cite{Friedman-mw,Curio:1998bva}.
Starting with the $E_8 \times E_8$ heterotic theory, the bundle
decomposes as $V_1 \oplus V_2$,  and without loss of generality, the curvatures split as \begin{equation}
\frac{1}{30}  {\rm Tr}\; F_i^2 = \eta_i \wedge \omega_0 + \zeta_i \,,
\;\;\;\;\; i = 1, 2
\label{eq:f2-decomposition}
\end{equation}
where $\eta_i, \zeta_i$ are (pullbacks of) 2-forms and 4-forms on
$B_2$ and $\omega_0$ is Poincar\'{e} dual to the zero-section of the elliptic fibration.  The heterotic Bianchi
identity \cite{GSW} gives $\eta_1 + \eta_2 = 12c_1 (B_2)$.  Thus, it is possible to parameterize a solution as
\begin{equation}
\eta_{1, 2} = 6c_1 (B_2)\pm T' \,, \;\;\;\;\;
(E_8 \times E_8)
\label{eq:eta}
\end{equation}
where $T'$ is a $\{ 1, 1 \}$ form on $B_2$. Next, returning to the F-theory geometry described above in \eref{eq:m}, the canonical class of ${\mathcal B}_3$ is determined by adjunction to be
\begin{equation}
-K_3 = 2 \Sigma_- -K_2 + T \,,
\label{eq:g}
\end{equation}
By studying the $4$-dimensional effective theories of these dual heterotic/F-theory compactifications it is straightforward to determine that the defining $\{1,1\}$ forms $T, T'$ in $B_2$ are in fact identical: $T=T'$ \cite{Friedman-mw, Grimm:2012yq}. The $\{1,1\}$-form $T$ is referred to as the ``twist" (of the $\mathbb{P}^1$-fibration) and is the crucial defining data of the simplest classes of heterotic/F-theory dual pairs.

\subsection{The spectral cover construction}
To explicitly match the degrees of freedom -- including the geometric moduli -- of a heterotic/F-theory dual pair, it is necessary to modify our description of the slope-stable holomorphic vector bundles arising as part of the heterotic background. A powerful tool to this end is the description of vector bundles known as the ``spectral cover construction\footnote{More generally, the ``cameral" cover construction \cite{donagi,Donagi:1998vw}.}" \cite{Friedman-mw,Donagi:1998vw,donagi,Ron_Elliptic}. In the simplest cases it is possible to form a $1-1$, onto map from a suitable\footnote{Here suitability is rigorously defined via the concept of ``regularity" \cite{Friedman:1997ih,Bershadsky:1997zv}.} slope-stable, holomorphic, rank $N$ vector bundle $\pi: V \to X_3$ to a pair $\left({\mathcal S}, {\mathcal L}_{{\mathcal S}} \right)$ (referred to as the  ``spectral data") where ${\mathcal S}$ is a smooth divisor in $X_3$ (forming an $N$-fold cover of the base $B_2$ and referred to as the ``spectral cover") and ${\mathcal L}_{\mathcal S}$ is a line bundle\footnote{More generally, a rank $1$ sheaf. For interesting physical examples where this distinction is crucial see \cite{Donagi:1995am,Aspinwall:1998he,Anderson:2013rka}.} over ${\mathcal S}$. 

The spectral cover construction has been used extensively in heterotic theories to construct rank $N$ bundles with structure group $SU(N)$ or $Sp(2N)$ that are slope-stable in some region of K\"ahler moduli space. As shown in \cite{Friedman-mw}, the class of the divisor ${\mathcal S}$ is given by
\begin{equation}\label{spec_class}
[{\mathcal S}]=N[\sigma]+\pi^*(\eta)
\end{equation}
where $\sigma$ is the zero section of $\pi: X_3 \to B_2$ and $\eta$ is
defined as in \eref{eq:f2-decomposition} and \eref{eq:eta}. 

It is helpful to once again describe the elliptically fibered heterotic threefold in Weierstrass form:
\begin{equation}\label{het_weir}
{\hat Y}^2={\hat X}^2 +f(u){\hat X}{\hat Z}^4 + g(u){\hat Z}^6
\end{equation}
where $\{{\hat X}, {\hat Y}, {\hat Z} \}$ are coordinates on the
elliptic fiber (described as a degree six hypersurface in
$\mathbb{P}_{123}$) and $\{u\}$ are coordinates on the base $B_2$. Here ${\hat
  Z}=0$ defines the section $\sigma$. For $SU(N)$ bundles, the 
{\it spectral cover}, ${\mathcal S}$, can
be represented as the zero set of the polynomial
\begin{equation}\label{speccov}
s=a_0 {\hat Z}^N+a_2 {\hat X}{\hat Z}^{N-2}+a_3 {\hat Y}{\hat Z}^{N-3} +\ldots
\end{equation}
ending in $a_N {\hat X}^{\frac{N}{2}}$ for $N$ even and $a_N {\hat
  X}^{\frac{N-3}{2}}{\hat Y}$ 
for $N$ odd
\cite{Friedman-mw}. The polynomials $a_i$ are sections of line bundles over the base
$B_2$ 
\begin{equation}
a_i \in H^0(B_2, K_{B_2}^{\otimes i} \otimes {\mathcal O}(\eta)) \,,
\end{equation}

In order for the spectral cover to be an actual algebraic surface in
$X_3$ (a necessary condition for the associated vector bundle to be Mumford slope-stability) it is necessary that ${\mathcal S}$ be an effective class in $H_4(X_3, \mathbb{Z})$. There is a further condition -- that the spectral cover must be {\it indecomposable} -- that must be imposed in order for the spectral cover bundle $V$ to be slope stable. It can be seen that ${\mathcal S}$ is indecomposable if $\eta$
is base-point free ({\it i.e.,} has no base locus in a Zariski-type decomposition and $\eta - Nc_1(B_2)$ is effective (see \cite{Donagi:2004ia} for example)).

All that remains to fully determine the holomorphic bundle $V$ is the data of the rank $1$ sheaf, ${\mathcal L}_{\mathcal S}$. As described in \cite{Friedman-mw}, given the projection $\pi_{\mathcal{S}}: {\mathcal S} \to B_2$, the Grothendieck-Riemann-Roch theorem \cite{hartshorne1977algebraic} indicates that
\beq
{\pi_{\mathcal S}}_{*}\left(e^{c_1({\mathcal L}_{\mathcal S})}Td({\mathcal S}) \right)= ch(\pi_*(V))Td(B_2)
\eeq
At the level of the first Chern class this yields
\beq
{\pi_{\mathcal S}}_{*} \left(c_1({\mathcal L}_{\mathcal S})+ \frac{1}{2}c_1({\mathcal S}) \right)=\frac{N}{2}c_1(B_2)+c_1(V)
\eeq
At this point, the condition that $c_1(V)=0$ (necessary for our choice of $SU(N)$ bundle $V\to X_3$) fixes $c_1({\mathcal L}_{\mathcal S}) \in H^{1,1}({\mathcal S}) \cap H^2({\mathcal S}, \mathbb{Z})$ up to a class $\gamma \in ker({\pi_{\mathcal S}}_*)$. Since $\pi_{\mathcal S}$ is an $N$-sheeted cover of $B_2$, ${\pi_{\mathcal S}}_{*}{\pi_{\mathcal S}}^{*}(c_1(B_2))=Nc_1(B_2)$ and hence
\beq\label{ls_general}
c_1({\mathcal L}_{\mathcal S})=\frac{N\sigma + \eta +c_1(B_2)}{2} + \gamma
\eeq
with
\beq
{\pi_{\mathcal S}}_{*}(\gamma)=0
\eeq

Here we are faced with the generally difficult problem of determining $\gamma$. We will return to this in the next section, but for now we simply review the observations made in \cite{Friedman-mw}: $c_1({\mathcal L}_{\mathcal S})$ must be an integral $(1,1)$-class on ${\mathcal S}$. For the cases of interest, such classes may be scarce since it can be verified that frequently $h^{2,0}({\mathcal S}) \neq 0$. As a result, the only obvious $(1,1)$-classes on ${\mathcal S}$ are those inherited from $X_3$, namely the restriction of the zero section of the elliptic fibration, $\sigma$, and pullbacks $\pi_{\mathcal S}^{*}(\beta)$ of integral $(1,1)$ classes on $B_2$.

Since ${\pi_{\mathcal{S}}}_{*} \sigma |_{{\mathcal S}}=\eta -Nc_1(B)$ one finds \cite{Friedman-mw} that a description of $\gamma \in ker({\pi_{\mathcal{S}}}_{*})$ in this ``obvious" basis is 
\beq\label{gammadef}
\gamma=\lambda(N \sigma|_{{\mathcal S}}-{\pi}_{\mathcal{S}}^{*}(\eta- N c_1(B))
\eeq
where $\lambda$ must be either integer or half integer according to
\begin{equation}\label{lambda_n}
    \lambda= 
\begin{cases}
    m+\frac{1}{2},& \text{if }~N ~\text{is odd}\\
    m,              & \text{if}~ N~\text{is even}
\end{cases}
\end{equation}
When $N$ is even it is clear that this {\bf integrality condition} imposes
\begin{equation}\label{mod2_cond}
\eta = c_1(B_2)~\rm{mod}~2
\end{equation}
where ``mod $2$'' indicates that $\eta$ and $c_1(B_2)$ differ only by an even element of $H^2(B_2, \mathbb{Z})$.
This leads to the form most commonly assumed in the literature \cite{Friedman-mw}:
\begin{equation}\label{l_top}
c_1(L_{\mathcal S})=N\left(\frac{1}{2}+\lambda \right)\sigma + \left(\frac{1}{2}- \lambda \right)\pi_{S}^{*}\eta + \left(\frac{1}{2}+N\lambda \right) \pi^{*}_{S} c_{1}(B_2)
\end{equation}

Having fully specified the topology of the spectral cover, it is possible to infer the full topology of $V$ itself. The Chern classes of a spectral cover bundle $V$, specified by $\eta$ and the integers $n$ and $\lambda$ is \cite{Friedman-mw,Friedman:1997ih,Curio:1998vu,Curio:2011eu}
\begin{align}
& c_1(V)=0 \\
& c_2(V)=\eta \sigma - \frac{N^3-N}{24}c_1(B_2)^2+\frac{N}{2}\left(\lambda^2-\frac{1}{4}\right)\eta \cdot \left(\eta-Nc_1(B_2) \right) \\
& c_3(V)=2\lambda \sigma \eta \cdot \left(\eta-Nc_1(B_2) \right)
\label{c3spec}
\end{align}
Note that since $c_1(V)=0$, ${\rm Ind}(V)=ch_3(V)= \frac{1}{2}c_3(V)$. 

The spectral cover construction provides a powerful tool in explicitly matching the geometric moduli of heterotic/F-theory dual pairs. For the details of the duality map and the necessary stable degeneration limit, we refer the reader to the classic references \cite{Friedman-mw, Curio:1998bva} and conclude here with only a rough hint in Table \ref{dualtable} of how the degrees of freedom associated to $(\mathcal{S}, \mathcal{L}_{\mathcal{S}})$ correspond to the moduli of a Calabi-Yau fourfold in F-theory.
\begin{table}[ht]
\begin{align}
&\text{Het/Bundle} & &\text{Het/Spec. Cov.}& & \text{F-theory}&  \nonumber \\ \hline
&H^1(End_0(V))& & H^{2,0}(\mathcal{S})\sim Def(\mathcal{S})& &H^{3,1}(\tilde{Y}_{4}) & \nonumber \\
& & &H^{1,0}(\mathcal{S})\sim Pic_0(\mathcal{S})& &H^{2,1}(\tilde{Y}_{4}) & \nonumber \\
& &&H^{1,1}(\mathcal{S})\sim \text{Discrete data of}~\mathcal{L}_{\mathcal{S}} & & H^{2,2}(\tilde{Y}_{4},\mathbb{Z}) & \nonumber
\end{align}
\caption{A rough, schematic matching of the heterotic vector bundle moduli, encoded as spectral data $(\mathcal{S}, \mathcal{L}_{\mathcal S})$, and geometric moduli of the (resolved) F-theory fourfold in the stable degeneration limit \cite{Friedman-mw, Curio:1998bva}.}\label{dualtable}
\end{table}
In later investigations, we will further compare the structure of an $SU(2)$ spectral cover with its F-theory dual consisting of a generically singular fourfold with $E_7$ symmetry. 

\section{A database of Heterotic/F-theory dual pairs}
In \cite{Anderson:2014gla} a systematic algorithm was laid out for constructing heterotic/F-theory dual pairs in which ${\mathcal B}_3$ (the base of the elliptically fibered fourfold geometry) is constructed as a $\mathbb{P}^1$ bundle over $B_2$. To illustrate the methods of construction, the complete dataset of Calabi-Yau fourfolds with smooth heterotic duals and \emph{toric} twofold bases were enumerated. This consisted of $4962$ Calabi-Yau fourfolds, dual to heterotic threefold/bundle geometry. Of these, $947$ were found to be generically singular with an $E_7$ symmetry (in at least one heterotic $E_8$ factor, equivalently $F$-theory coordinate patch). In the heterotic theory the $E_7$ gauge symmetry is realized by the commutant structure within $E_8$, via an $SU(2)$ vector bundle over the dual Calabi-Yau threefold. These rank $2$ vector bundles provide one of the simplest windows into the generic properties of the bundle moduli space ${\mathcal M}_{\omega}(c(V))$. Because of the fact that these $E_7$ symmetries are un-Higgsable  -- that is the fourfolds are generically singular for all values of the complex structure moduli, the results of Table \ref{corr_top} indicate that for this choice of $\eta$ the moduli space of stable sheaves contains only $SU(2)$ bundles.

Since the heterotic/F-theory duality map is most clearly understood in the case that the heterotic bundles can be described via spectral covers, it is natural to ask whether we can use this formalism to explicitly match the full degrees of freedom in dual $E_7$ effective theories described above.

As described in \cite{Anderson:2014gla}, the three conditions on the defining topological data, $\eta$, for consistent spectral covers are
\begin{itemize}
\item $\eta$ effective
\item $\eta$ base-point-free within $B_2$
\item $\eta=c_1(B_2)$ mod $2$
\end{itemize}

In \cite{Anderson:2014gla}, it was explored how these conditions compare to those arising in defining good Calabi-Yau fourfold backgrounds for F-theory. It can be shown that the first of these conditions is true for all $K3$-fibered fourfolds arising as F-theory backgrounds. Moreover, it can be shown that if the second condition is violated for a fourfold with a generic $E_7$ singularity, then the Calabi-Yau manifold is too singular to admit a K\"ahler resolution. To that point, the geometric consistency conditions on an F-theory fourfold and an $SU(2)$ heterotic spectral cover bundle are identical. However, as we will see, at the final condition, this agreement appears to end.

The condition $\eta=c_1(B_2)$ mod $2$ is required for the integrality of ${\mathcal L}_S$ in \eref{mod2_cond}. However, a direct construction of the dataset in \cite{Anderson:2014gla} shows immediately that this is violated for most fourfolds with generic $E_7$ symmetries -- in fact, $897$ of the $947$! How then are we to make sense of these dual pairs?

One obvious resolution to the puzzle could occur if none of the $897$ moduli spaces of $SU(2)$ bundles could admit any bundle built via the spectral cover construction. While possible, this seems unlikely from experience of how generic spectral cover bundles appear to be in known moduli spaces \cite{Bershadsky:1997zv}. Another possible answer is that the integrality condition placed on $c_1({\mathcal L}_{\mathcal S})$ in \eref{mod2_cond} may be artificially restrictive. This will clearly be the case whenever the Picard number of ${\mathcal S}$ is greater than $1+h^{1,1}(B_2)$ as assumed by \cite{Friedman-mw}. 

One class of examples in which the Picard group of ${\mathcal S}$ is larger than the generic case was outlined in \cite{Curio:2011eu}. There, it was pointed out that if the matter curve $a_2=0$ in \eref{su2spec} (in the class $[\eta -2c_1(B_2)]$) is reducible in $B_2$, its components may in fact pull back to distinct, new divisors in ${\mathcal S}$. That is, if the curve $\bar{\eta} \in [\eta -2c_1(B_2)]$ can be written as $\bar{\eta}=D+ D'~~\subset B_2$,
then its pullback can be described as
\beq
\pi^*_{\mathcal S}(\bar{\eta})={\mathcal D} + \mathcal{D'}
\eeq
and even if $D, D'$ are well-understood divisors in $B_2$, the class ${\mathcal D}$ in ${\mathcal S}$ may not be a simple linear combination of the divisors $\sigma|_{\mathcal S}$ and $\pi^{*}_{\mathcal S} (\phi)$ (with $\phi$ an effective curve class in $B_2$) assumed in the generic formula \eref{ls_general}. In \cite{Anderson_stringmath} we explored whether or not this observation could alleviate the disparity of the mysterious $897$ $E_7$ theories found in \cite{Anderson:2014gla}. While a handful of the examples found over Hirzebruch bases could be resolved by this mechanism, the majority of them remained unexplained \cite{Anderson_stringmath}. To really resolve this puzzle and decide whether or not these geometries consist of valid heterotic/F-theory dual pairs, it is necessary to go further and attempt to study the integrality condition in detail. We turn to this now in the context a simple example of an $SU(2)$ bundle defined over $\pi_h: X_3 \to \mathbb{P}^2$.

\section{A case study: bounding the Picard number $\rho({\mathcal S})$}\label{picard}
To begin, it is useful to summarize the discussion of the previous sections in the context of an $SU(2)$ spectral cover. To fully specify the $SU(2)$ gauge bundle appearing in the heterotic compactification, it is not enough to choose a spectral cover of the form given in \eref{spec_class} and \eref{speccov}, we must also fully describe the line bundle, ${\mathcal L}_{\mathcal S}$ over $\mathcal{S}$. A priori, we can describe the 1st Chern class of ${\mathcal L}_{\mathcal S}$  via \eref{ls_general} as
\beq
c_1({\mathcal L}_{\mathcal S})=\frac{N\sigma + \eta +c_1(B_2)}{2} + \gamma
\eeq
where ${\pi_{\mathcal S}}_{*} (\gamma)=0$. By the construction of ${\mathcal S} \subset X_3$ there are $1+h^{1,1}(B_2)$ natural integral $(1,1)$-classes on ${\mathcal S}$ (consisting of the restriction of $\sigma$, the section of the elliptic fibration, and the pullback of classes from the base). Using these as basis (and ignoring any other possibilities for $\gamma$) the integrality condition given in \eref{gammadef} and \eref{mod2_cond} were obtained in \cite{Friedman-mw}. However, recent work on the F-theory side of the duality \cite{Anderson:2014gla} indicates that this integrality condition appears to be violated in the vast majority of known examples ($897$ of $947$ generic $E_7$ models enumerated in \cite{Anderson:2014gla}) we must now ask whether or not it is possible to derive a more general integrality condition for $c_1({\mathcal L}_{\mathcal S})$? To accomplish this, ${\mathcal L}_{\mathcal S}$ must be expressed in a complete basis. We are thus led to the following question:
\begin{quest}
For a general surface ${\mathcal S} \subset X_3$ (as described above) which is a ramified, $N$-sheeted cover of $B_2$ in the class $[N\sigma + {\pi_h}^*(\eta)]$ what is the rank of the Picard group of $\mathcal{S}$?
\end{quest}
As illustrated in the next Section, generically $h^{1,0}(\mathcal{S})=0$ \cite{Friedman-mw} and $\mathcal{S}$ is a surface of general type. Unfortunately, determining the Picard number of such complex surfaces is a notoriously difficult problem (see \cite{poonen,schutt,beauville} and references therein for some recent advances). To begin, it is enough to consider ways to bound the Picard number $\rho({\mathcal S})$ as an important first step. 

Let us briefly recall a few standard definitions regarding the Picard group (see \cite{hartshorne1977algebraic,miranda} for example). To define divisors (and hence line bundles), one begins with the exponential sequence
\beq
0 \to \mathbb{Z} \stackrel{i}{\longrightarrow} \mathcal{O} \stackrel{exp}{\longrightarrow} \mathcal{O}^{*} \to 0
\eeq
where the map $i$ is an inclusion and $exp$ is the exponential map. With vanishing $Pic_0(\mathcal{S})$ (i.e. with $h^{1,0}(\mathcal{S})=h^1(\mathcal{S}, \mathcal{O})=0$ there are no continuous degrees of freedom in the Picard group), the associated long exact sequence in cohomology takes the form
\beq
0 \to H^1(\mathcal{S}, \mathcal{O}^{*} ) \to H^2(\mathcal{S}, \mathbb{Z}) \to H^2(\mathcal{S}, \mathcal{O})
\eeq
The image of $H^1(\mathcal{S}, \mathcal{O}^{*} )$ (modulo torsion) in $H^2(\mathcal{S}, \mathbb{Z})$ parametrizes the Neron-Severi group, $NS(\mathcal{S})$, of the surface and its rank is the Picard number (i.e. $\rho(\mathcal{S})$, the number of discrete parameters which we can use to construct $\mathcal{L}_{\mathcal{S}}$). The Picard group is given by the kernel of the map from $H^2(\mathcal{S}, \mathbb{Z})$ to $H^2(\mathcal{S}, \mathcal{O})=H^{0,2}$. The Hodge decomposition and Lefschetz' theorem \cite{hartshorne1977algebraic} demonstrate that it is also zero in $H^{2,0}$ and hence must be a subset of $H^{1,1}$:
\beq
NS(S) \simeq H^2(\mathcal{S}, \mathbb{Z}) \cap H^{1,1}(\mathcal{S})
\eeq
Stated simply, divisors on ${\mathcal S}$ are determined by how the complex subspace $H^{1,1}$ of $H^2(\mathcal{S}, \mathbb{C})$ intersects the discrete subgroup $H^{2}(\mathcal{S}, \mathbb{Z})$. For surfaces with vanishing geometric genus ($p_{g}=h^{0,2}=0$) this is a trivial identification, but few tools exist to address the general case with $p_g \neq 0$. To begin, it should be observed that there is at least a bound:
\beq
\rho(\mathcal{S}) \leq h^{1,1}(\mathcal{S})
\eeq

Since in the present work we are focused on the case of $2$-sheeted spectral covers and the mysterious $E_7$ cases described in the previous section, here we will try to make a first step towards answering this question. We will consider a simple example appearing in \cite{Anderson:2014gla}, with $\pi_{\mathcal S}: \mathcal{S} \to \mathbb{P}^2$. As we will see, even here determining the full Neron-Severi group is a non-trivial problem in algebraic geometry and for this brief work, we content ourselves with simply bounding the Picard number, $\rho({\mathcal S})$ as described above.

\subsection{A double cover of $\mathbb{P}^2$}
In an explicit example we can explore in detail the possible form of the spectral line bundle, $\mathcal{L}_{\mathcal S}$. We consider here a $2$-sheeted spectral cover, $\mathcal{S}$, and one of the simplest examples arising in the dataset of \cite{Anderson:2014gla}. Let $\pi: X_3 \to \mathbb{P}^2$ be a Calabi-Yau threefold described via the generic (smooth) Weierstrass model over $\mathbb{P}^2$:
\begin{equation}\label{het_weir2}
{\hat Y}^2={\hat X}^2 +f(u){\hat X}{\hat Z}^4 + g(u){\hat Z}^6
\end{equation}
where $u_i$ ($i=1,2,3$) are homogeneous coordinates of $\mathbb{P}^2$ and 
\beq
f \in H^0(\mathbb{P}^2, \mathcal{O}(12H))~~~,~~~g \in H^0(\mathbb{P}^2, \mathcal{O}(18H))
\eeq
where $H$ is the hyperplane divisor in $\mathbb{P}^2$. This Weierstrass model can be realized as hypersurface inside a toric variety. In a language more familiar to physicists, this threefold also be written via a GLSM-style charge matrix (see \cite{Blumenhagen:2010pv} for example):
\begin{table}[ht]
  \centering
  \begin{tabular}{c|cccccc}
     & $\hat{Y}$  & $\hat{X}$  & $\hat{Z}$  & $u_1$  & $u_2$ & $u_3$        \\ 
    \hline 
    6  & 3 & 2 & 1 & 0 & 0 & 0 \\
   0 & 0  & 0  & -3 & 1 & 1 & 1\\
  \end{tabular}
  \label{tablepolyzeros}
\end{table} \\
The hodge numbers of this threefold are well-known to be $h^{1,1}=2$, $h^{2,1}=272$. Furthermore, a basis of divisors on $X_3$ is given by $D_1, D_2$ where $D_2=\pi^{*}(H)$ is the pullback of the hyperplane in $\mathbb{P}^2$ and $D_1$ is related to the elliptic fiber such that the class of the zero section ($\hat{Z}=0$) is given in this basis as $\sigma=D_1-3D_2$. The tangent bundle of $X_3$ is described via adjunction as
\beq\label{tang}
0 \to TX_3 \to T{\mathcal A}|_{X_3} \to \mathcal{O}(6D_1)|_{X_3} \to 0
\eeq
where $T{\mathcal A}$ denotes the tangent sheaf of the toric ambient space. This in turn is defined by an Euler sequence \cite{hartshorne1977algebraic}:
\beq
0 \to {\mathcal O}^{\oplus 2} \to {\mathcal O}(3D_1) \oplus {\mathcal O}(2D_1) \oplus {\mathcal O}(D_1 -2 D_2) \oplus {\mathcal O}(D_2)^{\oplus 3} \to T{\mathcal A} \to 0
\eeq

For this geometry we specify vector bundles and a dual F-theory geometry by making a choice of twist as in Section \ref{het_fth}, eq.\eref{eq:eta}. Here we select
\beq
T=10H
\eeq
In the heterotic theory this leads to an $SU(2)$ bundle $V \to X_3$ with 
\beq
\eta=6c_1(\mathbb{P}^2) -T=8H
\eeq
In the dual F-theory geometry this corresponds to a Calabi-Yau fourfold with {\emph generic} $E_7$ singularity \cite{Anderson:2014gla}. From \eref{spec_class}, the spectral cover is in the class $[{\mathcal S}]=[2\sigma +8\pi^*(H)]$ which in the basis given above corresponds to a section of the line bundle $N_{\mathcal{S}}=\mathcal{O}(2D_1+2D_2)$. Explicitly  $\mathcal{S}$ is given by \eref{speccov} as the zero locus of
\beq\label{su2spec}
a_{0}{\hat Z}^2+a_{2}{\hat X}=0
\eeq
with $a_0 \in H^0(\mathbb{P}^2, {\mathcal O}(8H))$ and $a_2 \in H^0(\mathbb{P}^2, {\mathcal O}(2H))$. Let us now take a closer look at ${\mathcal S}$. The complex, K\"ahler surface is a ramified double cover of $\mathbb{P}^2$ and we can directly compute its three independent Hodge numbers 
\beq
h^{2,0}(\mathcal{S}), ~~h^{1,0}(\mathcal{S}),~~h^{1,1}(\mathcal{S})
\eeq
To explicitly determine these numbers, we can once again make use of an adjunction formula, this time for ${\mathcal S}$ itself as a hypersurface inside $X_3$:
\beq\label{shype}
0 \to T{\mathcal S} \to TX_3|_{\mathcal S} \to \mathcal{O}(2D_1 +2D_2)|_{\mathcal S} \to 0
\eeq
Furthermore, to determine the cohomology of vector bundles restricted to ${\mathcal S}$, the Koszul sequence for hypersurfaces
\beq\label{koszul_s}
0 \to \mathcal{O}_{X_3}(-2D_1 -2D_2) \to \mathcal{O}_{X_3} \to \mathcal{O}_{\mathcal S} \to 0
\eeq
and its associated long exact sequence in cohomology plays a useful role (see \cite{Anderson:2008ex} for a review). In the case at hand, all the relevant cohomology groups on $X_3$ can be determined by considering the defining sequences \eref{shype}, \eref{tang} and \eref{koszul_s} and line bundle cohomology on $X_3$. For this geometry we employed the techniques of \cite{Blumenhagen:2010pv} to compute line bundle cohomology on $X_3$ (as implemented in \cite{CohomCalg}).

To begin, we note that $h^{2,0}({\mathcal S})=H^0({\mathcal S}, \mathcal{O}(2D_1+2D_2)|_{\mathcal S})$. Twisting \eref{koszul_s} by $\mathcal{O}(2D_1+2D_2)$ we obtain
\beq
0 \to \mathcal{O}_{X_3} \to \mathcal{O}_{X_3}(2D_1+2D_2) \to \mathcal{O}_{\mathcal S}(2D_1+2D_2) \to 0
\eeq
The associated long exact sequence in cohomology leads to
\beq
H^0({\mathcal S}, \mathcal{O}(2D_1+2D_2)|_{\mathcal S})=H^0(X_3, \mathcal{O}(2D_1+2D_2))/\mathbb{C}
\eeq
Which can be directly calculated to yield $h^0({\mathcal S}, \mathcal{O}(2D_1+2D_2)|_{\mathcal S})=h^0(X_3, \mathcal{O}(2D_1+2D_2))-1=51-1$. This provides the first of three independent hodge numbers (the geometric genus):
\beq
h^{2,0}(\mathcal{S})=50
\eeq
Note that this is expected via the description of ${\mathcal S}$ in \eref{su2spec}. By inspection of that formula it can be noted that there are $51$ degrees of freedom in the coefficients $a_0,a_2$ over $\mathbb{P}^2$. Subtracting $1$ for the overall scale, we see that this agrees with the expectation of the embedding moduli of ${\mathcal S} \subset X_3$.

Next, note that $h^{1,0}=h^{1}(\mathcal{S}, \mathcal{O}_{\mathcal S})$ (the ``irregularity" of the surface). Here the long exact sequence in cohomology associated to \eref{koszul_s} yields
\beq
h^{1,0}({\mathcal S})=0
\eeq
Finally, to determine $h^{1,1}(\mathcal{S})$, consider the dual sequence
\beq\label{dualseq}
0 \to  \mathcal{O}(-2D_1 -2D_2)|_{\mathcal S} \to {TX_3}^{\vee}|_{\mathcal S} \to T{\mathcal S}^{\vee} \to 0 
\eeq
To evaluate this it should first be noted that the Koszul sequence for $\mathcal{O}(-2D_1 -2D_2)$ produces the following short exact sequence
\beq
0 \to \mathcal{O}_{X_3}(-4D_1 -4D_2) \to \mathcal{O}_{X_3}(-2D_1 -2D_2) \to {\mathcal O}_{\mathcal S}(-2D_1 -2D_2) \to 0 
\eeq
 and from the associated sequence in cohomology
\begin{align}
& h^0(\mathcal{S},{\mathcal O}_{\mathcal S}(-2D_1 -2D_2))= h^1(\mathcal{S},{\mathcal O}_{\mathcal S}(-2D_1 -2D_2))=0  \label{data1}\\
& h^2(\mathcal{S},{\mathcal O}_{\mathcal S}(-2D_1 -2D_2))=219 \nonumber
\end{align}
This gives the full cohomology of the first term bundle in \eref{dualseq}. But what is $H^*(\mathcal{S}, {TX_3}^{\vee}|_{\mathcal S} )$? The last necessary pieces can be obtained by considering \eref{koszul_s} twisted by $TX_3^{\vee}$:
\beq
0 \to TX_3 \otimes \mathcal{O}_{X_3}(-2D_1-2D_2) \to TX_{3}^{\vee} \to TX_{3}^{\vee}|_{\mathcal S} \to 0
\eeq
Here the long exact sequence in cohomology produces
\begin{align}
& h^0({\mathcal S}, TX_{3}^{\vee}|_{\mathcal S})=0 \label{data2} \\
& h^1({\mathcal S}, TX_{3}^{\vee}|_{\mathcal S})=h^1(X_3, TX^{\vee})+ dim(ker(\phi))=2+dim(ker(\phi)) \nonumber \\
&h^2({\mathcal S}, TX_{3}^{\vee}|_{\mathcal S})=dim(coker(\phi)) \nonumber \\
& \phi: H^2(X, TX_3 \otimes \mathcal{O}_{X_3}(-2D_1-2D_2) ) \to H^2(X, TX_3^{\vee}) \nonumber
\end{align}
Since $h^2(X, TX_3 \otimes \mathcal{O}_{X_3}(-2D_1-2D_2))=393$ and $h^2(X, TX_3^{\vee})=272$, it follows that $dim(ker(\phi))=121+m$ for some $m \geq 0$, and $dim(coker(\phi))=m$ by exactness.  In fact, for generic choices of spectral cover in \eref{su2spec}, we expect the induced map $\phi$ to be surjective and $h^1({\mathcal S}, TX_{3}^{\vee}|_{\mathcal S})=123$.

With this in hand, we are now in a position to put the pieces together to determine $H^1(\mathcal{S}, T\mathcal{S}^{\vee})$. Using \eref{data1} and \eref{data2}, and returning to the long exact sequence in cohomology associated to \eref{dualseq} gives the following long exact sequence:
\beq\label{lastlong}
0 \to H^2(\mathcal{S},TX_{3}^{\vee}|_{\mathcal S}) \to H^{1,1}(\mathcal{S}) \to H^2(\mathcal{S}, \mathcal{O}_{\mathcal{S}}(-2D_1 -2D_2)) \to H^2(\mathcal{S},TX_{3}^{\vee}|_{\mathcal S}) \to 0
\eeq
It is helpful to note that $h^2(\mathcal{S}, T\mathcal{S}^{\vee})=h^{1,0}=0$, and the alternating sum of the dimensions in \eref{lastlong} leads at last to 
\beq
h^{1,1}(\mathcal{S})=(123+m) +(219-m)=342
\eeq
Thus, in summary we have determined that ${\mathcal S}$ is a complex surface with $h^{1,0}=0$, $h^{2,0}=50$ and $h^{1,1}=342$. It follows that the Euler number of ${\mathcal S}$ is $e=2+2p_g+h^{1,1} -4h^{1,0}=444$ (with $e=c_2(T{\mathcal S})$) and the holomorphic Euler characteristic is $\chi=51$ (leading to $K_{\mathcal{S}}^{2}=168$). According to Kodaira's classification, $\mathcal{S}$ is a surface of general type (Kodaira dimension 2).

\vspace{15pt}

Taking a step back, one can now ask what we have learned from the this example? The first observation is that in this case
\beq
2 \leq \rho(\mathcal{S}) \leq 342
\eeq
where the lower bound arises from concrete construction of divisors \cite{Friedman-mw} and the upper bound is obtained from $h^{1,1}$ as described in the previous Subsection. It should be noted here that there are in principle hugely more parameters in the spectral data than are commonly assumed in the physics literature. While the full computation of $\rho(\mathcal{S})$ is beyond the scope of the present work, tools exist to analyze the intersection structure of curves in ${\mathcal S}$ and can be used to further constrain $\rho(\mathcal{S})$ in many cases. We hope to explore this in future work.  For the moment, in the example above, we expect that $H^2(\mathcal{S}, \mathbb{Z}) \cap H^{1,1}(\mathcal{S})$ will generically be large. Indeed, despite the fact that $p_g=50$, $h^{1,1}$ is sufficiently big that contrary to the expectations of \cite{Friedman-mw}, it may be that the Picard number $\rho(\mathcal {S})$ is considerably above its minimum value of $2$. In this case, there are certainly more general choices available for the line bundle, $\mathcal{L}_{\mathcal{S}}$, and the integrality condition in \eref{mod2_cond} is manifestly incorrect and too restrictive. 

To proceed further with this explicit example, it might be possible to consider the branch locus of the two-sheeted cover in detail. Such an analysis was undertaken in \cite{persson} for certain double covers of $\mathbb{P}^2$. There for special choices of topology, the resolution of singularities in the branch curve led to concrete descriptions of the Neron-Severi group of the double cover (which was in fact maximal in those cases). It would be interesting in the future to explore the application of these techniques to heterotic spectral covers.

Finally it should be noted that as $\mathcal{S}$ is varied within the $50$-parameter family given in \eref{su2spec}, the Picard number can surely change. While difficult to compute, these special, higher codimensional ``Noether-Lefschetz Loci" \cite{kim, green} may be especially significant for the underlying physics, determining for example, where the complex structure moduli of the dual F-theory geometry, $Y_4$ are stabilized by $G$-flux \cite{Donagi:2009ra}.

To conclude, the example above was provided as a simple illustration of the fact that the integrality condition for spectral cover bundles given in \eref{mod2_cond} may be too restrictive in many cases. Furthermore, it serves to highlight the interesting and frequently difficult geometric questions that arise in fully determining the geometry of dual heterotic/F-theory pairs. As a final comment on the mysterious $897$ $E_7$ examples highlighted in \cite{Anderson:2014gla}, the arguments presented above indicate to us that in fact there is more to understand about integrality conditions in spectral covers and that this may provide a resolution to the seeming discrepancy in all the exotic heterotic/F-theory pairs. We hope in future work to build upon the simple examples considered here and to fully compute the Picard group of ${\mathcal S}$ systematically in the full dataset. By addressing these remaining geometric puzzles we hope it will be possible to complete the program laid out in \cite{Anderson:2014gla} and fully enumerate all consistent heterotic/F-theory dual pairs. 

\section*{Acknowledments}
The author would like to thank J. Gray and W. Taylor for useful discussions. The work of L.A. is supported by NSF Grant PHY-$1417337$.

\bibliographystyle{amsalpha}

\end{document}